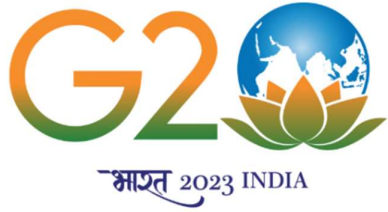 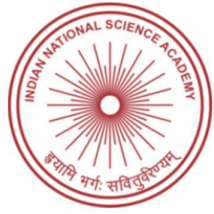 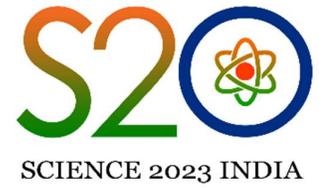

# Policy Brief

## LATEST DEVELOPMENTS AND OPPORTUNITIES IN SKY SURVEYS

Discussed and drafted during S20 Policy Webinar on Astroinformatics for Sustainable Development held on 6-7 July 2023

Contributors: Anthony Brown, Federica Bianco, Varun Bhalerao, Shri Kulkarni, Jeffery Cooke, David Reitze, Pranav Sharma, Ashish Mahabal

INDIAN NATIONAL SCIENCE ACADEMY – CENTRE FOR SCIENCE POLICY AND RESEARCH

**Introduction**
Studying the Universe, and the phenomena within it, requires astronomers to observe specific objects in detail or observe many objects simultaneously over large swaths of the sky using purpose-built wide-field instruments. The latter programs are referred to here as 'sky surveys'. As these surveys are time-, facility-, and data-intensive, they are performed either as single well-designed surveys, typically collecting a large amount of data and probing deep into the Universe, or by scanning the sky many times in search of objects that vary over time, or move that exhibit transient phenomena such as explosions, or outbursts. Some surveys combine both approaches.

Sky surveys have been a crucial tool in advancing our understanding of the Universe. They have helped us discover new astronomical objects, the origin of the elements, dark matter and dark energy, the accelerated expansion of the universe, and gravitational waves. They have helped us study the distribution of neutral and ionized matter in the Universe and test our theories about the origin and evolution of galaxies, stars, and planets. The last few decades have seen an explosion in the number and scope of sky surveys, both ground-based and space-based. This growth has led to a wealth of data that has enabled us to make significant advances in many areas of astronomy, and help understand the physics of the universe. The diagram below gives an impression of the growth of sky surveys in terms of the amount of data collected per night of observation.

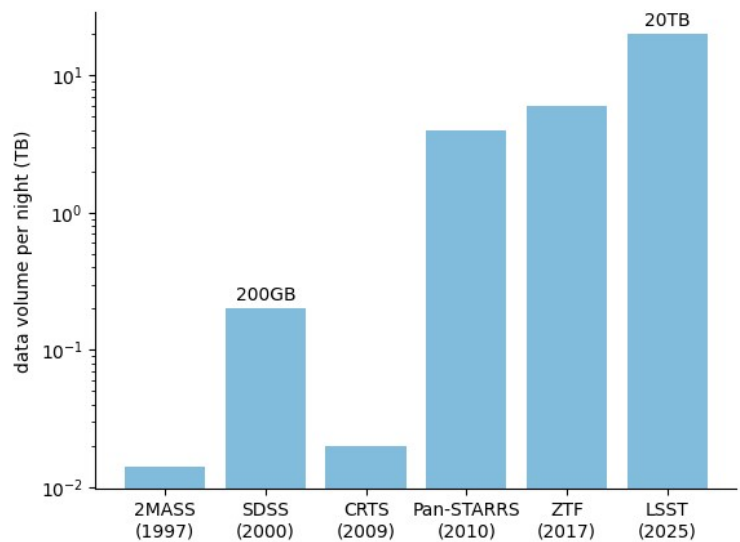

In this policy brief, we will explore recent advancements and potential avenues in sky surveys, encompassing both terrestrial and spaceborne initiatives. We will also examine how these developments may impact the field of international astronomical research.

**Current State of Sky Surveys**
The current state of sky surveys is characterized by a large number of ongoing and planned surveys, covering a wide range of wavelengths, messengers, and timescales. Some of the most notable surveys in the optical and infrared include the Two Micron All Sky Survey (2MASS), the Sloan Digital Sky Survey (SDSS), the Catalina Real-Time Transient Survey (CRTS), the Panoramic Survey Telescope and Rapid Response System (Pan-STARRS), the Dark Energy Survey (DES), the Zwicky Transient Survey (ZTF), the (upcoming) Vera C. Rubin Observatory Legacy Survey of Space and Time (LSST), and, in space, the European Space Agency's Gaia and Euclid, and NASA's Roman Space Telescope missions to name a few.

At wavelengths other than ultraviolet/optical/infrared sky surveys have been and are being conducted both at lower and higher energies, respectively in the radio and X-ray/γ-ray/cosmic-ray domains. Notable examples include the gravitational wave surveys (LIGO-Virgo-KAGRA), the Low Frequency Array (LOFAR) and the future Square Kilometre Array (SKA) and its current pathfinders, e.g., ASKAP and MWA, in the radio domain. High energy sky survey facilities include the Cherenkov Telescope Array (CTA), the space-based Fermi Gamma-ray Burst Monitor (GBM), and particle detectors such as the IceCube Neutrino Observatory, the Pierre Auger Observatory, the High-Altitude Water Cherenkov Observatory (HAWC), and the KM3Net neutrino telescope. Both the SKA and CTA offer examples of cross-continental collaboration on large sky surveys, with both surveys operating telescopes in multiple continents.

In parallel dedicated spectroscopic surveys of the sky have also increased in number, including the Radial Velocity Experiment (RAVE), the Large Sky Area Multi-Object Fibre Spectroscopic Telescope (LAMOST) surveys, the GALactic Archaeology with HERMES (GALAH) survey, the Apache Point Observatory Galactic Evolution Experiment (APOGEE) surveys, and the upcoming surveys WHT Enhanced Area Velocity Explorer (WEAVE), 4-meter Multi-Object Spectroscopic Telescope (MOST), and SDSS-V.

These surveys have produced, are producing, or will produce massive amounts of data, posing significant challenges in terms of data storage, processing, and analysis. At the same time these surveys offer tremendous opportunities for scientific research across the world, including in countries without the resources to initiate their own sky surveys.

**Important Developments**
The developments of sky surveys over the past decade have focused on improving the depth, coverage, and resolution. One of the notable developments is the combination of data from surveys conducted in different wavelength regimes to provide a more complete picture of astronomical objects. Another development is the use of time-domain surveys that monitor the sky for changes in brightness or position, enabling the study of transient phenomena such as supernovae, gamma-ray bursts and all kinds of Galactic variables and transients. For example, CHIME covers the full sky essentially everyday, as do Fermi and IceCube. In the optical ZTF covers the northern sky every few nights, and soon the LSST will image the entire southern sky to greater depths every few days. This will vastly improve our knowledge of the dynamical night sky.

**Opportunities for Global Astronomy Research**
Sky surveys provide a wealth of opportunities for global astronomy research. With large sky footprints and broad temporal and wavelength coverage by nature, these surveys generate data that can be used to address a wide range of scientific questions, including the nature of dark matter and dark energy, the formation and evolution of galaxies, the study of exoplanets and their atmospheres, stellar physics, and the search for extraterrestrial intelligence.

Sky surveys, i.e. surveys designed to map the sky as opposed to study specifically pre-identified objects, have been conducted since the end of the 19th century. The use of data from surveys conducted at different wavelength regimes or from surveys conducted at different epochs provide a powerful astronomical research tool. For example, the ultra-precise positions of celestial objects measured by the Gaia mission can be used to tie together photographic surveys from a century ago with modern digital sky surveys, thus enabling studies over long-time baselines. These facilitate a much better understanding of small bodies in the solar system, among other science goals. The search for Near-Earth objects has been a motivation for many surveys, and their cataloguing is important to secure humanity's continued existence. Similarly, the combination of the Hipparcos and Gaia space missions (25 years difference) has greatly improved the efficiency at finding wide orbit exoplanets around nearby stars. In the future, similar opportunities can be exploited in the domain of spectroscopic surveys, provided the data is archived properly and kept easily accessible.

To fully exploit the opportunities provided by sky surveys, it is essential to ensure that the data is archived and accessible to researchers worldwide. This requires the continued development of open data policies and the establishment of data-sharing platforms that allow researchers to access and analyse the data. This has been the goal of the Virtual Observatories which provide an excellent practical example of the required policies and platforms.

**Recommendations**[1]

1. Collaboration: Encourage increased collaboration between different sky survey teams and institutions to improve data sharing and reduce duplication of efforts.
2. Standardization: Standardizing survey data formats (e.g., units, features, and other aspects) will greatly help optimize scientific output and help promote cross-discipline and sub-discipline work.
3. Data Quality: Establish guidelines and best practices for data quality control, documentation, provenance, and archiving to ensure the integrity and reliability of survey data. This will facilitate future analysis and comparison with other data sets. In today's world the concept of data integrity should be expanded to include robustness against contamination of data archives with "fake data" generated by modern AI systems.
4. Cross-Disciplinary Research: Encourage cross-disciplinary research by providing opportunities for astronomers to collaborate with researchers from other fields, such as computer science and statistics, but also fields such as geography and medicine, and sub-disciplines (other wavelengths and messengers), to explore new methods of data analysis and visualization.
5. Long Term Data Preservation: Develop the tools, protocols, and platforms needed to ensure that sky survey data is preserved in usable form. This demands an ongoing effort to evolve the preservation strategies along with changes in storage technology.
6. Accessibility: Develop accessible and user-friendly platforms and tools for accessing and analysing survey data, such as online data portals and interactive visualizations. Specifically, the development of data analysis platforms that are close to the data should be stimulated so that scientists without the institutional means can still access the processing power needed to analyse large data sets (across time). Survey teams should collaborate in the development of platforms with shared or similar design, to avoid the need for scientists to learn different working modalities, an additional workload that would particularly impact scientists in under resourced countries or institutions where research time and support may be limited.
7. Education and Outreach: Promote the use of survey data in education and outreach activities to engage the public and inspire the next generation of astronomers and data scientists and scientists of all types. In this context it is important that astronomical survey data are made publicly available (which is already the case for many surveys) for use by non-scientists, educators, and outreach professionals. This offers a unique opportunity to engage communities in countries with less direct access to astronomical survey facilities.

---

[1] The following recommendations are in line with recent major strategic reports:
   1. *Pathways to Discovery in Astronomy and Astrophysics for the 2020s*, Decadal Survey on Astronomy and Astrophysics 2020 (2021)
   2. The Astronet Science Vision and Infrastructure Roadmap 2022–2035 (2023)

8. Funding:
    a) Secure funding for the continued operation (i.e., running the telescopes and instruments and carrying out the data processing) and maintenance of sky surveys, including support for upgrades and the development of new survey instruments and technologies.
    b) Secure funding to make sky survey data publicly accessible. This requires investment in long term archiving capacity (storage, curation) and in the accessibility tools mentioned above.
    c) Secure funding to support community studies and the training and education of the scientific community on the use of the survey data and platforms from the start. The training should include the skills needed to analyse large data sets with machine learning and artificial intelligence tools. All desiring communities should have the opportunity to learn how to use survey data products for research and be ready for the survey when it starts. This recommendation enables and supports research inclusion, which should be built into the design of the survey and with dedicated resources, so as to not to end up marginalizing under resourced communities, or disadvantaged communities (e.g. the blind and visually impaired) compromising equity and losing the discovery potential they bring to the table.
9. Recognition and Rewards: The astronomical community now expects survey data to be publicly available in a timely fashion which means that data rights for surveys can only be exercised for a limited period. This can be demotivating for (young) researchers who want to contribute to surveys but also want to benefit from proprietary periods to create and publish scientific results, which are still an important component in evaluations for stable jobs in academia. The academic community, with support from governments, should look into ways of creating career paths where one is rewarded and promoted for one's contribution to making large surveys possible. This requires the development of evaluation criteria that balance individual versus team efforts and reward contributions to open science.

**Conclusion**

Sky surveys are an essential tool for advancing our understanding of the Universe. The current state of sky surveys is characterized by many ongoing and planned surveys, producing massive amounts of data. Recent developments in sky surveys have focused on improving the depth, coverage, and resolution (both spatial and temporal) of surveys. Sky surveys provide a wealth of opportunities for global astronomy research, and it is essential to ensure that the data is maintained and accessible to researchers worldwide for the long term.

**S20 Co-Chair**: Ashutosh Sharma, Indian National Science Academy
**INSA S20 Coordination Chair:** Narinder Mehra, Indian National Science Academy

**Contributors**
Anthony Brown, Leiden University, Netherlands
Federica Bianco, University of Delaware, USA
Varun Bhalerao, Indian Institute of Technology – Mumbai, India
Shri Kulkarni, California Institute of Technology, USA
Jeffery Cooke, Swinburne University of Technology, Australia
David Reitze, California Institute of Technology, USA
Pranav Sharma, Indian National Science Academy, India
Ashish Mahabal, California Institute of Technology, USA